\documentclass[prl,letterpaper,aps,10pt,superscriptaddress,twocolumn,floatfix,showpacs]{revtex4-2} 
\usepackage{graphicx}
\usepackage{amsmath,amssymb,amsfonts}
\usepackage{graphicx,braket,xcolor,subfigure,upgreek}
\usepackage[colorlinks, linkcolor=blue, citecolor=blue, urlcolor=blue, breaklinks=true]{hyperref}
\usepackage{microtype}
\usepackage{bbm}
\usepackage{color}
\usepackage[english]{babel}
\usepackage[toc,page]{appendix}

\newcommand{\fref}[1]{Fig.~\ref{#1}}

\graphicspath{{figures/}}

\begin{document}

\title{Cavity Sub- and Superradiance Enhanced Ramsey Spectroscopy}

\author{Christoph Hotter}
\affiliation{Institut f\"ur Theoretische Physik, Universit\"at Innsbruck, Technikerstr. 21a, A-6020 Innsbruck, Austria}
\author{Laurin Ostermann}
\affiliation{Institut f\"ur Theoretische Physik, Universit\"at Innsbruck, Technikerstr. 21a, A-6020 Innsbruck, Austria}
\author{Helmut Ritsch}
\affiliation{Institut f\"ur Theoretische Physik, Universit\"at Innsbruck, Technikerstr. 21a, A-6020 Innsbruck, Austria}
\date{\today}

\begin{abstract}
Ramsey spectroscopy in large, dense ensembles of ultra-cold atoms trapped in optical lattices suffers from dipole-dipole interaction induced shifts and collective superradiance limiting its precision and accuracy. We propose a novel geometry implementing fast signal readout with minimal heating for large atom numbers at lower densities via an optical cavity operated in the weak single atom but strong collective coupling regime. The key idea is controlled collective transverse $\pi/2$-excitation of the atoms to prepare a macroscopic collective spin protected from cavity superradiance. This requires that the two halves of the atomic ensemble are coupled to the cavity mode with opposite phase, which is naturally realized for a homogeneously filled volume covering odd and even sites of the cavity mode along the cavity axis. The origin of the superior precision can be traced back to destructive interference among sub-ensembles in the complex nonlinear collective atom field dynamics. In the same configuration we find surprising regular self-pulsing of the cavity output for suitable continuous illumination. Our simulations for large atom numbers employing a cumulant expansion are qualitatively confirmed by a full quantum treatment of smaller ensembles.
\end{abstract}


\maketitle

\section{Introduction}
Precise measurement of the energy difference  of atomic transitions and their corresponding frequencies has been at the heart of atomic physics for more than a century. Nowadays, laser spectroscopy features incredible accuracy and precision of up to $20$ digits \cite{giunta201920}. These advances allow for precise tests of fundamental physics, from general relativity and astrophysics to effects of super-symmetry in strong interactions with nuclei or even the spectral properties of antimatter \cite{ludlow2015optical, poli2013optical}. Yet, there is still a strong desire to push these limits even further, while, in parallel, one also aims at technologically simpler, faster and more robust methods to reach a desired precision.

Ramsey spectroscopy, employing a carefully engineered sequence of time delayed short phase-coherent pulses, can be used to resonantly excite a particle to a long-lived and thus energetically very narrow state. It is one of the most commonly used and well proven methods in precision spectroscopy. Applied to large ensembles of ultra-cold atoms trapped in a magic wavelength optical lattice it constitutes the current state of the art \cite{norcia2019seconds}. In order to realize precise measurements fast, a large number of particles needs to be interrogated. They should be confined to a small volume in order to ensure a homogeneous environment. Here, the remaining weak inter-particle interactions are a source of inaccuracy and the emerging fast superradiant collective decay limits the maximal measurement time and precision \cite{Kramer2016optimized}.

The measurement readout requires a precise determination of the fraction of excited atoms for a given probing frequency. In principle, one has to measure the induced fluorescence after the Ramsey sequence. However, for very long lived clock states this is too slow, inefficient and imprecise in order to be practical and more elaborate procedures are necessary. Hence, besides methods relying on additional fast transitions \cite{dehmelt1982monoion}, using cavity enhanced detection of dispersive shifts or collective decay has been suggested \cite{lodewyck2009nondestructive, ludlow2015optical}. However, while cavity superradiance is fast and precise for counting excitations \cite{black2005ondemand, Zhang2012Collective}, the presence of the cavity perturbs the free evolution between the Ramsey pulses, which again limits measurement precision.
\begin{figure}[t]
\centering
\includegraphics[width=0.85\columnwidth]{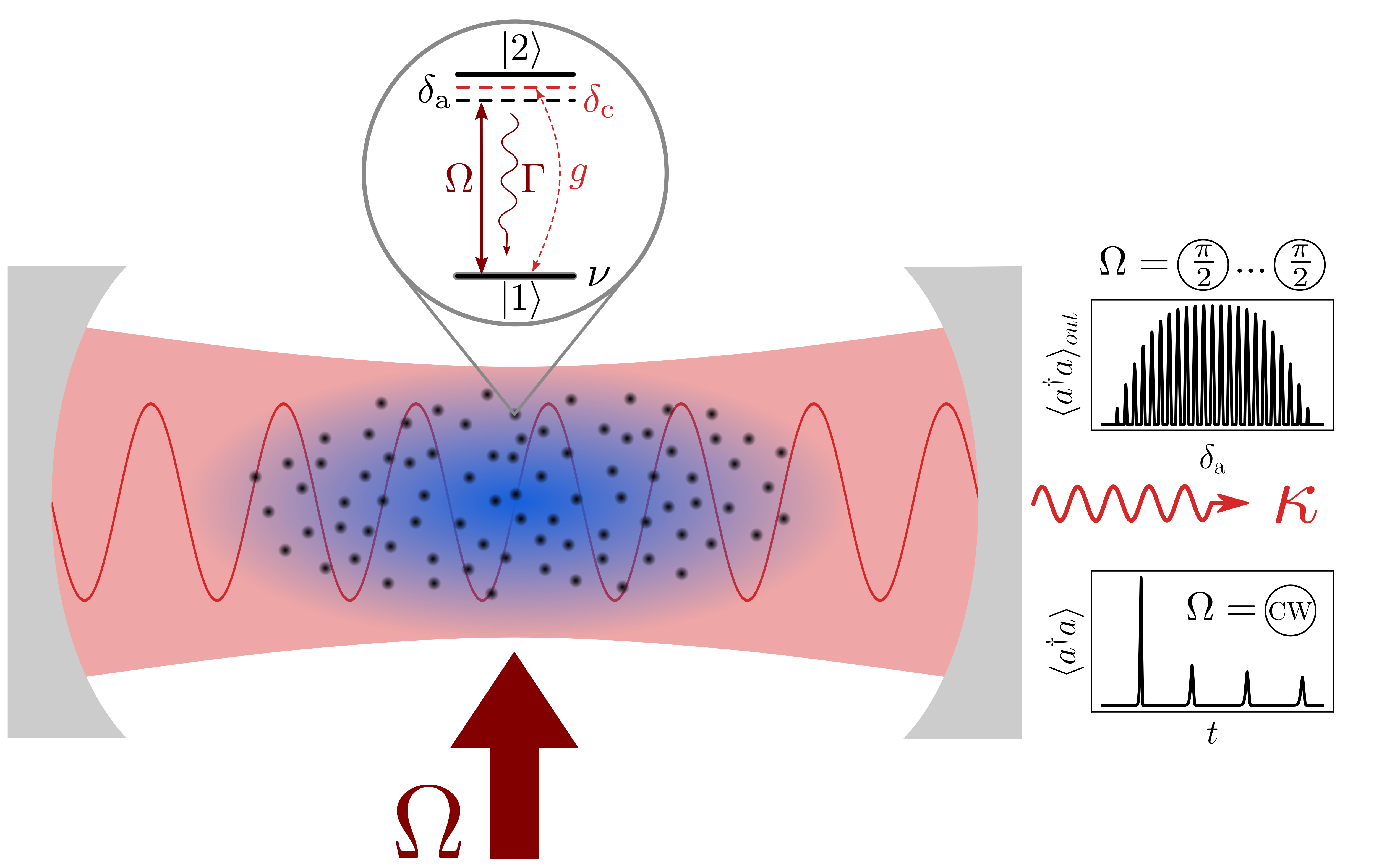}
\caption{\emph{Cavity Enhanced Spectroscopy Setup.} We consider a homogeneous, dilute ensemble of narrow line two-level atoms at random, but fixed positions in a standing wave optical resonator coherently driven by a transverse plane wave laser. We assume a weak single atom but strong collective coupling regime, i.e.\ $ \kappa, \,\sum_j \, g_j^2 / \kappa \gg \Gamma \gg g_j^2 / \kappa $. The Cavity Ramsey fringes and the photon number self-pulsing are indicated as cavity output signals for the respective drive laser operations.}
\label{fig:model}
\end{figure}

In this work we propose an improved implementation of cavity enhanced Ramsey spectroscopy combining fast collective readout with high precision. The central idea is to address special collective excited states of the atomic ensemble, which decouple from the cavity during the free evolution phase \cite{ostermann2013protected,ostermann2014protected}.  The final readout is still performed via the superradiant manifold allowing for fast detection of the ensemble. Surprisingly, despite sounding challenging and complicated, there is a very natural and simple possibility to implement this scheme for a large but dilute ensemble within a standing wave cavity.

\emph{Model.} -- %
We consider $N$ two-level atoms with a narrow transition at frequency $\omega_\mathrm{a}$ coupled to a single mode cavity. The atoms are coherently driven with a detuning between the laser and the atomic transition frequency of $\delta_\mathrm{a} = \omega_\mathrm{l} - \omega_\mathrm{a}$, the corresponding Rabi frequency is denoted by $\Omega$. The cavity is detuned by $\delta_\mathrm{c} = \omega_\mathrm{l} - \omega_\mathrm{c}$ from the laser and we have an atom-cavity coupling of $g_j$ for the $j$-th atom. The system is depicted in~\fref{fig:model}. Its Hamiltonian  in the rotating frame of the pump laser reads
\begin{equation} \begin{aligned}
H &= - \delta_\mathrm{c} a^\dagger a + \sum_{j=1}^{N} \Big[- \delta_\mathrm{a} \sigma_j^{22}  \\
& + g_j (a^\dagger \sigma_j^{12} + a \sigma_j^{21}) + \frac{\Omega}{2} (\sigma_j^{21} + \sigma_j^{12}) \Big] ,
\label{eq:hamiltonian}
\end{aligned} \end{equation}
with the cavity photon creation (annihilation) operator $a^\dagger$ ($a$) and the atomic transition operator $\sigma_j^{kl} = \ket{k}_j \bra{l}_j$ for the $j$-th atom. The coherent interaction is accompanied by dissipative processes, which are accounted for by the Liouvillian $\mathcal{L} \left[ \rho \right]$ in the master equation,
\begin{equation}
\dot{\rho} = i \left[ \rho, H \right] + \mathcal{L} \left[ \rho \right].
\label{eq:master}
\end{equation}
In the Born-Markov approximation~\cite{gardiner2004quantum} we can write the Liouvillian in Lindblad form as
\begin{equation}
\mathcal{L} \left[ \rho \right] = \sum_i R_i \left( 2J_i \rho J_i^\dagger - J_i^\dagger  J_i \rho - \rho J_i^\dagger  J_i \right),
\label{eq:liouvillian}
\end{equation}
with the jump operator $J_i$ and its corresponding rate $R_i$ for the $i$-th dissipative process. We list these processes in Tbl.~\ref{tab:dissipative}, including cavity photon losses as well as decay and dephasing of the atoms with rate $\Gamma$ and $\nu$, respectively.
\begin{table}[b]
\caption{\emph{Dissipative Processes.} The system features a damped cavity mode as well as atomic decay and dephasing.}
\begin{center}
\begin{tabular}{ lccl }
 \hline
 $i$ & $J_i$ & $R_i$ & Description \\
 \hline
 1 & $a$ & $\kappa$ & cavity photon losses \\
 2 & $\sigma_j^{12}$ & $\Gamma$ & decay from $\ket{2}_j$ to $\ket{1}_j$ \\
 3 & $\sigma_j^{22}$ & $\nu$ & dephasing of the $j$-th atom \\
\hline
\end{tabular}
\end{center}
\label{tab:dissipative}
\end{table}

As we are targeting narrow clock transitions for our atoms the system is operated in the bad and large volume cavity regime $\kappa \gg \Gamma$ with only a small single atom cooperativity $C_j = g_j^2/(\kappa \Gamma) \ll 1$ but a sufficiently large ensemble to enter the strong collective coupling regime $ NC = \sum_j C_j \gg 1$.  Typically, this parameter regime implies a very large atom number $N \gg 1$, which does not allow for a full quantum simulation, but we can very well treat this problem in a second order cumulant expansion~\cite{Kubo1962generalized, plankensteiner2021quantumcumulantsjl}. A comparison with a full quantum simulation for a small atomic ensemble is shown in the appendix~\cite{2nd_order_example}. For a qualitative illustration of our deliberations even first order cumulant equations, i.e. mean field, (see appendix) suffice for most calculations. Additionally, we neglect dipole-dipole interaction~\cite{ficek2002entangled}, as our atomic ensemble is sufficiently dilute such that neither dispersive, nor dissipative collective processes among the individual atoms will play a considerable role.

\emph{Collective Cavity mediated Super- and  subradiance.} -- %
In the following, for simplicity, we assume the atoms located close to cavity mode anti-nodes with half of the atoms at the maxima and half at the minima of the mode function along the cavity axis. Hence their respective effective coupling is well approximated by $+g$ and $-g$. As confirmed by more involved simulations, investigating various different distributions for the atom-field coupling, this simplification already captures the essential physics discussed below.
\begin{figure}[t]
\centering
\includegraphics[width=\columnwidth]{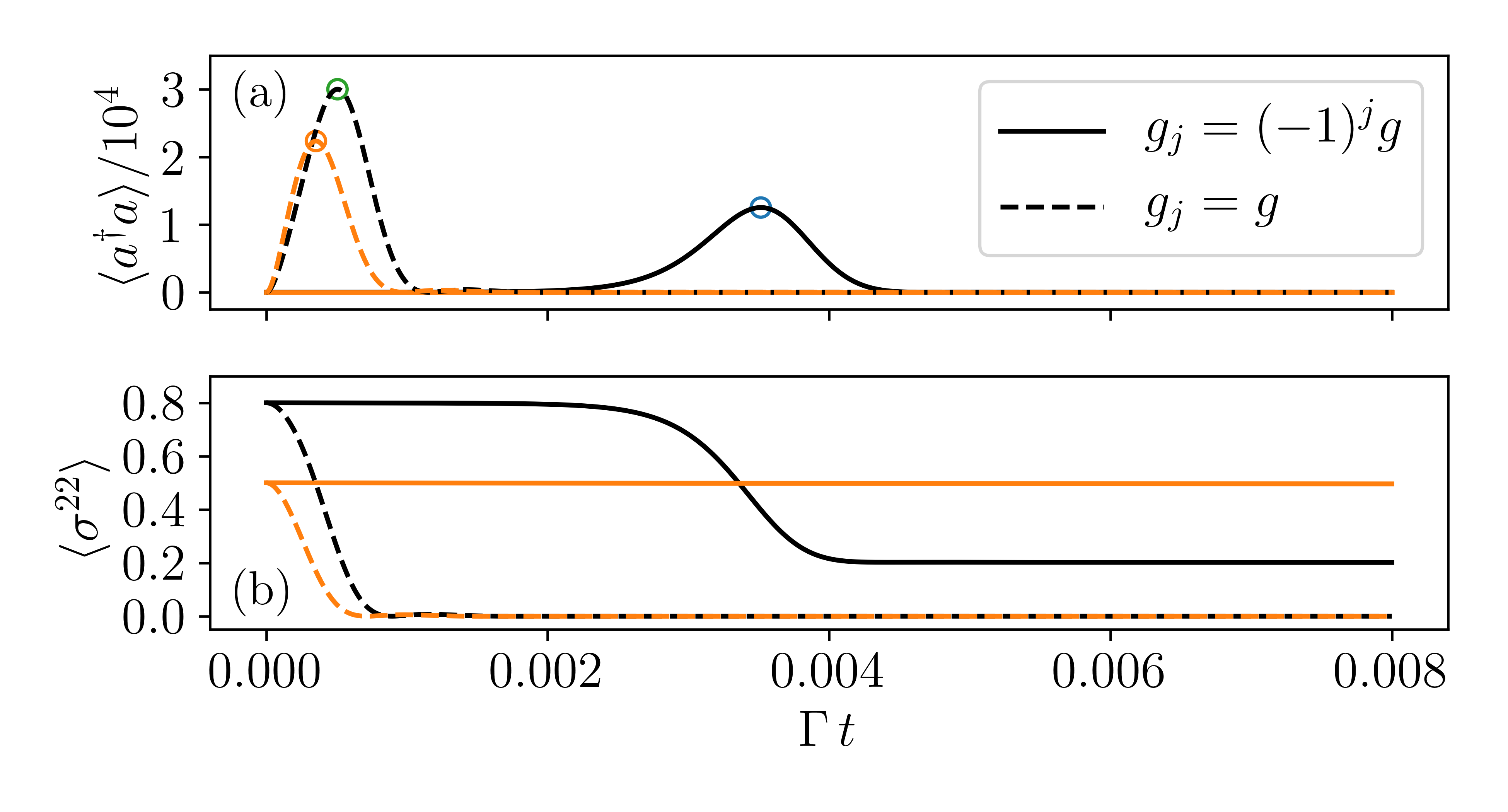}
\vspace{-0.6cm}
\caption{\emph{Cavity Mediated Collective Decay.} Time evolution of the intra-cavity photon number (a) and the single atom excited state population (b) after a short pulse excitation preparing each atom in the same coherent superposition. The black line represents an inverted ensemble with $\left \langle \sigma^{22} \right \rangle = 80\%$ excited state fraction ($\approx 3\pi/4$-pulse) and the orange line depicts an ensemble after a $\pi/2$-pulse to create $\left \langle \sigma^{22} \right \rangle = 50\%$. For uniform cavity coupling (dashed line) in both cases we see an immediate superradiant population decay to the ground state creating a photon pulse in the cavity. Cavity coupling with alternating signs (solid line) leads to a weaker and time delayed pulse for an inverted ensemble, while the atomic excitation is almost perfectly protected from cavity decay without inversion (solid line in (b)). We have assumed $N = 2 \cdot 10^5$ atoms with  $ g = 10\Gamma, \, \kappa = 10^4 \Gamma$  and $\delta_\mathrm{a} = \delta_\mathrm{c} = \nu = 0 $.}
\label{fig:superradiance}
\end{figure}

It is well established that inverting all atoms with a short $\pi$-pulse induces the emission of a delayed intense light pulse due to cavity enhanced superradiant decay~\cite{mlynek2014observation,norcia2016superradiance, norcia2018frequency, laske2019pulse, schaffer2020lasing, tang2021cavity}.  Synchronized stimulated emission in a cavity occurs even for a dilute ensemble, which does not exhibit free space superradiance. Figure~\ref{fig:superradiance}a shows typical trajectories for the corresponding time evolution of the intra-cavity photon number $\left \langle a^\dagger a \right \rangle$, when all atoms are initially coherently prepared at $\left \langle \sigma^{22} \right \rangle = 80\%$ (black line). Figure~\ref{fig:superradiance}b depicts the corresponding time evolution of the excited state population. For comparison we show the behavior for all atoms equally coupled to the cavity (dashed line, $g_j = g$) as well.

\begin{figure}[b]
\centering
\includegraphics[width=\columnwidth]{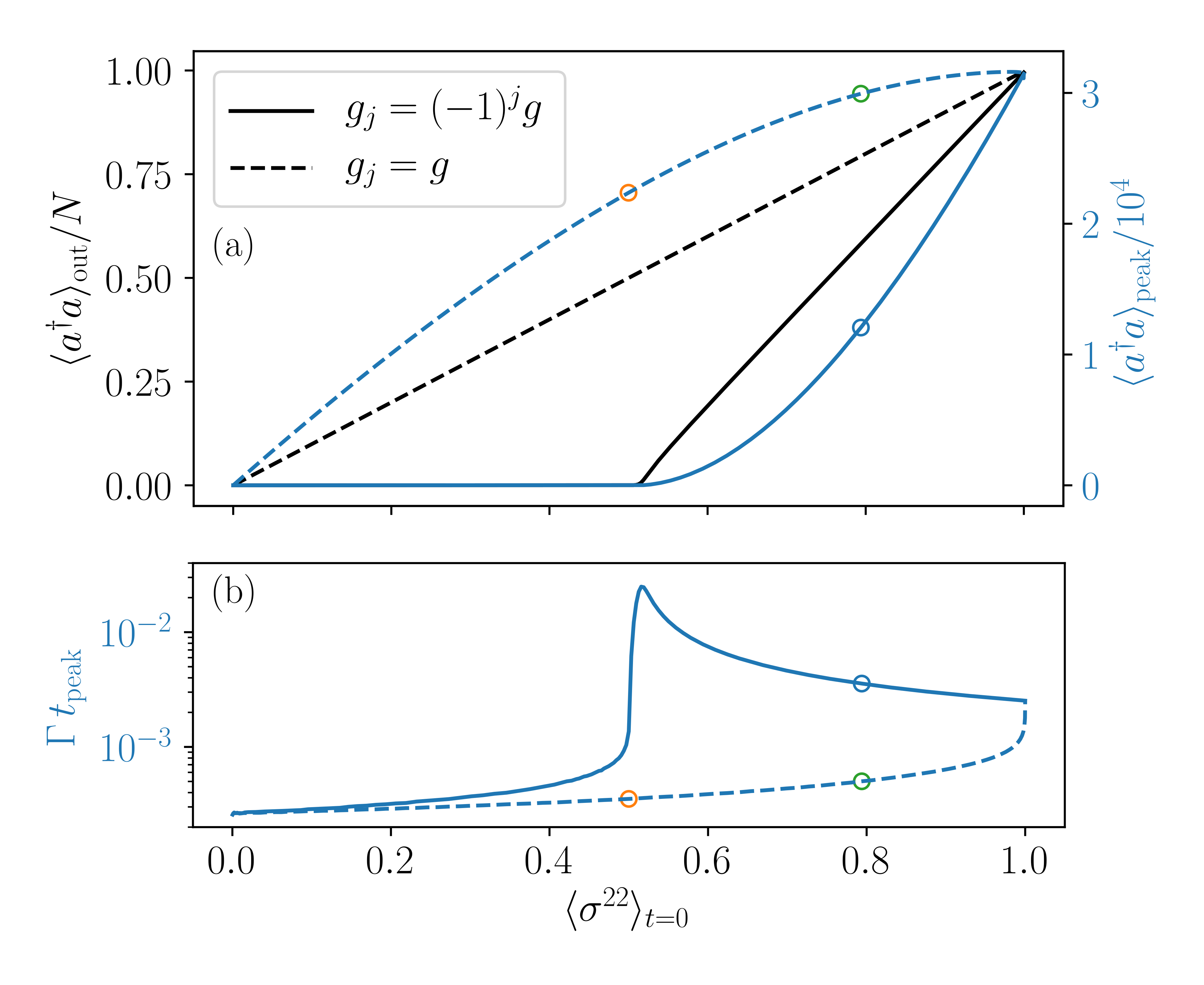}
\vspace{-0.6cm}
\caption{\emph{Cavity Sub- and Superradiance.} (a) Comparison of the integrated cavity-output photon number $\left \langle a^\dagger a \right \rangle_\mathrm{out} = \kappa \int \left \langle a^\dagger a \right \rangle \mathrm{d}t$ (black) and peak intra-cavity photon number (blue) as a function of the single atom excitation probability for alternating coupling (solid line) and uniform coupling (dashed line). Note the strong suppression of superradiant emission in the alternating coupling case as long as no net-inversion is created initially. (b) Delay time of the peak photon number. For alternating coupling the pulse appears earlier for higher inversion, whereas for equally coupled atoms the pulse delay time increases with growing excited state population. The circles indicate the operating parameters chosen for~\fref{fig:superradiance}a.}
\label{fig:pulse}
\end{figure}

However, for the system we consider ($g_j = (-1)^j g$), we will observe such pulsed emission for an ensemble of inverted atoms only. If the excited state population is below $50\%$ the atoms are not able to synchronize and thus will not emit a significant amount of photons into the cavity, see solid orange line in \fref{fig:superradiance}. Figure~\ref{fig:pulse}a shows the total number of emitted photons $\langle a^\dagger a \rangle_\mathrm{out}$ for different values of the coherently prepared initial excited state population. We can clearly observe that almost no photons leak through the cavity mirrors until the atoms are inverted. This is due to subradiant suppression as the photons emitted by the atoms coupled to the cavity with opposite $g$ will interfere destructively \cite{filipp2011preperation, reimann2015cavity}. For a fully inverted ensemble we obtain approximately one photon for each atom. However, for an initial inversion with $\left \langle \sigma^{22} \right \rangle < 1$ we only get $N \cdot (2 \langle \sigma^{22} \rangle - 1)$ photons. The solid lines in~\fref{fig:superradiance}b also indicate that the atoms retain a population of $1 - \left \langle \sigma^{22} \right \rangle$ after the photon emission into the cavity and subsequently experience free-space decay with the rate $\Gamma$ only. Again, the dashed line represents the case of all atoms coupling equally ($g_j = g$), where we see that even not fully inverted atoms will superradiantly emit photons  into the cavity. Additionally we plot the peak intra-cavity photon number (blue) illustrating the same behavior. Figure \ref{fig:pulse}b shows the delay time of the peak photon number as a function of the atomic excitation. For uniform coupled atoms (dashed line) a higher excitation leads to a later pulse. Whereas for alternating coupling (solid line) the delay time of the peak decreases for higher inversion, with a maximum at $\langle \sigma^{22} \rangle_{t=0} = 50\%$.

Note, that a crucial requirement for cavity subradiance is that the cumulative dipole moment of the atoms projected on the cavity mode $\sum_j g_j \langle\sigma^{12}_j \rangle$  vanishes~\cite{ostermann2013protected, ostermann2014protected}. Here, for simplicity, we chose the same laser excitation phase for all atoms, but a variable excitation phase works equally well as long as the overall relative phase disappears. Fortunately, this case is typically automatically realized for a random distribution of a sufficiently large dilute intra-cavity atomic ensemble. Note that this is also true for a ring-cavity featuring a continuous atom-cavity coupling phase along the cavity axis~\cite{ritsch2012cavitypotentials}.

\emph{Cavity Enhanced Ramsey Probing.} -- %
As argued above, transverse pumping with an overall vanishing phase of the atom-cavity coupling thus allows for a $\pi/2$-pulse excitation of the atomic ensemble without an immediate rapid superradiant decay through the cavity. This intriguing feature will be an essential ingredient for a new implementation of cavity assisted Ramsey spectroscopy as discussed below, when we  combine it with fast direct measurements of the number of excited atoms via the emitted cavity photons after the second Ramsey pulse. The crucial advantage of this scheme is that it can be very fast with no additional manipulation of the atoms needed for the read out, hence the signal is less perturbed. Furthermore the atoms are not significantly heated by this measurement and can therefore be reused, thus, the dead time between measurements can be significantly reduced. Another advantage in comparison with other non-demolition measurements for atomic clocks~\cite{lodewyck2009nondestructive, vallet2017noise, hobson2019cavity} is that the signal, i.e.\ the number of photons, scales linearly with the number of atoms. So, in principle, an arbitrarily large number of atoms can be employed, which drastically increases the signal to noise ratio.

\begin{figure}[t]
\centering
\includegraphics[width=\columnwidth]{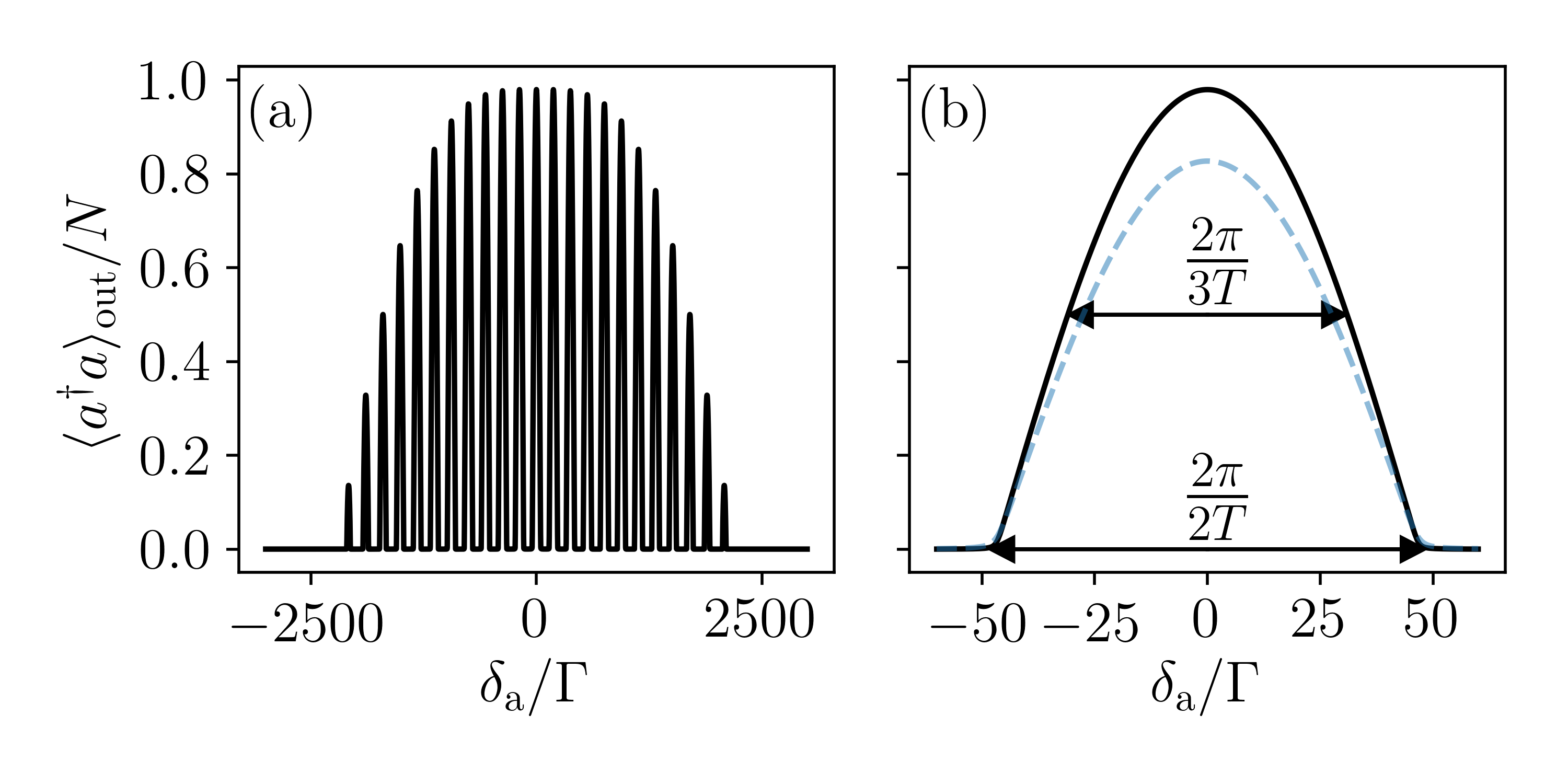}
\vspace{-0.6cm}
\caption{\emph{Cavity Ramsey Method.} (a) Fringes in the photon number obtained via the cavity Ramsey method with characteristic flat zero-photon regions. (b) Zoom-in on the central fringe. The FWHM ($\pi/T$) of an optimal independent atom Ramsey sequence with waiting time $T$ and the cavity Ramsey fringe ($\sim 2\pi/3T$) are highlighted. The parameters are $N = 2 \cdot 10^5, \, g = 10\Gamma, \, \kappa = 10^4 \Gamma, \, \delta_\mathrm{c} = \delta_\mathrm{a}, \, \Omega = 1000\Gamma$  and $\nu = 0$ (solid) or $\nu = 10\Gamma$ (dashed). The free evolution time between the two $\pi/2$-pulses is $T=\pi/100\Gamma$.}
\label{fig:ramsey}
\end{figure}

Figure~\ref{fig:ramsey}a shows the output signal of the cavity Ramsey method, the total number of photons leaking through the cavity mirrors as a function of the laser-atom detuning $\delta_\mathrm{a}$. Similar to the conventional Ramsey method, fringes appear~\cite{ramsey1950molecular, sortais2001cold, bize2005cold}. One striking difference, however, is that a non-inverted ensemble does not produce a signal, corresponding to the flat zero-photon regions. This narrows the FWHM of the cavity-Ramsey fringes slightly compared to the conventional Ramsey fringes (see~\fref{fig:ramsey}b). Including an atomic dephasing with $\nu = 10\Gamma$ (dashed line) merely weakens the signal, yet, the shape of the curve is essentially the same. Note, that by choosing $\delta_\mathrm{c} = \delta_\mathrm{a}$ we have implicitly assumed that the cavity is perfectly on resonance with the atomic transition. Therefore, one might wonder if a detuned cavity impairs the signal. But, since we operate deeply in the bad cavity regime, only shifts of the cavity resonance frequency on the order of $\kappa$ are important.

Overall, this means that enhancing the Ramsey spectroscopy by adding a cavity achieves the same (or even slightly improved) accuracy, but has the advantage of a convenient, fast, non-destructive measurement scaling linearly in the atom number, which can substantially reduce the measurement dead time. At this point we want to mention that the atoms need to be initially in the ground state for each subsequent measurement. As we saw in \fref{fig:superradiance}b a not fully inverted ensemble will retain some population in the excited state. This means one needs to bring these atoms back to the ground state. Unfortunately, this is not possible in a straight-forward way via a coherent drive on the clock transition only. However, there are other ways to achieve this, for example with an induced decay via another transition or by depleting the ground state population temporarily to some other level to enable the depopulation of the excited clock state.

Note that this cavity assisted Ramsey procedure will not work for $\pi/2$-pulse excitations through the cavity mirrors. The reason for this is that all atoms exhibit the same relative atom-cavity phase and will therefore already decay superradiantly after the first $\pi/2$-pulse.

\begin{figure*}[t]
\centering
\includegraphics[width=\textwidth]{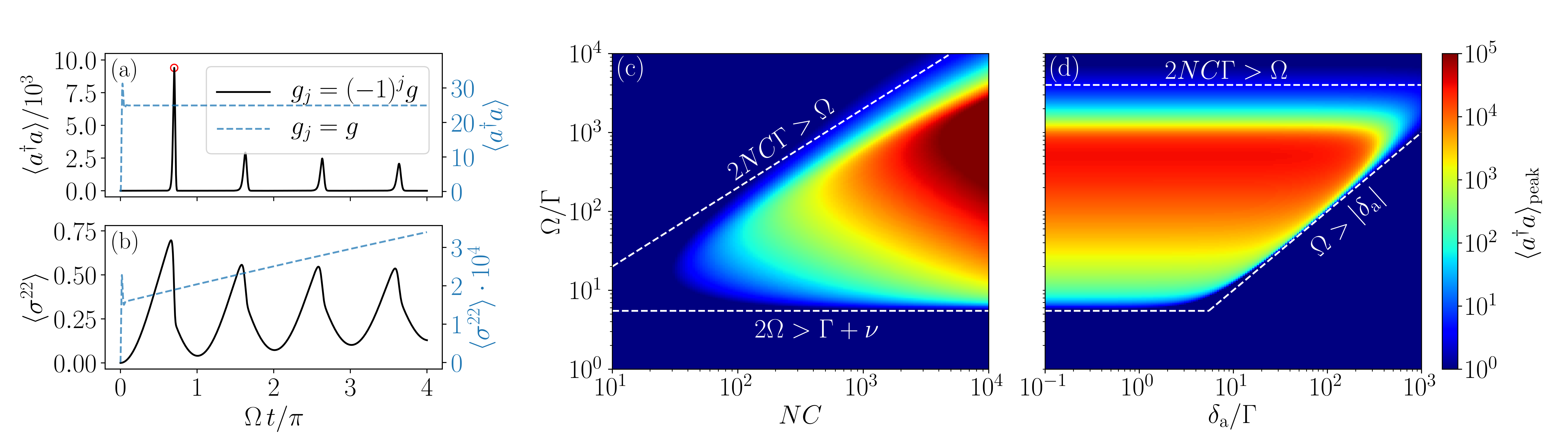}
\vspace{-0.6cm}
\caption{\emph{Self-Pulsing.} Time evolution of the cavity photon number (a) and excited state population (b) for a continuous drive, resulting in photon number self-pulsing. (c) and (d) Scans over $\Omega$, $NC$ and $\delta_\mathrm{a}$ for the peak photon number of the first pulse (red circle in (a)). The dashed white lines represent the threshold $2 N C \Gamma > \Omega > |\delta_\mathrm{a}|, (\Gamma + \nu) / 2$. The parameters when kept constant are $N = 2 \cdot 10^5, \, g = 10\Gamma, \, \kappa = 10^4 \Gamma, \, \delta_\mathrm{c} = \delta_\mathrm{a} = 0, \, \Omega = 100\Gamma$  and $\nu = 10$.}
\label{fig:dynamic_pulses}
\end{figure*}

\emph{Self-Pulsing under Continuous Illumination.} -- %
Continuously driving the atomic ensemble with a suitable Rabi frequency leads to striking self-pulsing of the system as shown in~\fref{fig:dynamic_pulses}a. Yet, the explanation for this initially surprising behavior is rather simple: as we have seen, the photon emission into the cavity for a non-inverted atomic ensemble with vanishing relative phase is strongly suppressed. Therefore, with atoms initially in the ground state, there is no significant cavity photon number at least until $t=\pi/2\Omega$ (see~\fref{fig:dynamic_pulses}b). But, as soon as a significant population inversion is achieved the ensemble emits a superradiant photon pulse into the cavity. Subsequently the excited state population is depleted below $50\%$ and the photon number quickly reduces to almost zero due to the very fast cavity photon decay. Since the laser is still on, the procedure starts over again and we obtain another pulse. As we can see in~\fref{fig:dynamic_pulses}a the peak photon number reduces from pulse to pulse, which can be traced back to decay and decoherence damping the Rabi oscillations. Additionally the timeevolution for all equally coupled atoms (dashed blue line, $g_j = g$) is plotted in (a) and (b). We see that in this case the self-pulsing does not occur and the cavity photon number reaches a steady state at $\langle a^\dagger a \rangle = \Omega^2/4g^2$ very quickly. The steady state value for the excited state population ($\langle \sigma^{22} \rangle \approx 1.7 \cdot 10^{-3}$) is reached much later. Note, that the system with vanishing relative phase also reaches a steady state with, surprisingly, the same photon number $\langle a^\dagger a \rangle = \Omega^2/4g^2$ but a much higher excited state population ($\langle \sigma^{22} \rangle \approx 0.25$).

As we need an excited state population of at least $50\%$ to observe the photon peaks, and the atoms basically do Rabi oscillations to reach this, there should be a lower bound for the Rabi frequency $\Omega$, depending on the laser-atom detuning $\delta_\mathrm{a}$, the atomic decay $\Gamma$ and the dephasing rate $\nu$. This is exactly what we see in~\fref{fig:dynamic_pulses}c and~\fref{fig:dynamic_pulses}d. The condition $\Omega > |\delta_\mathrm{a}| , \, (\Gamma + \nu)/2$ needs to be satisfied in order for the peak intra-cavity photon number of the first pulse to appear. Furthermore, the collective photon emission from the atoms into the cavity is determined by the frequency $N C \Gamma$. Thus, for the superradiant photon pulse to dominate over the coherent photon scattering from the drive, we need to ensure that $N C \Gamma > \Omega / 2$. This threshold is also shown in~\fref{fig:dynamic_pulses}c and \fref{fig:dynamic_pulses}d. Additionally we find that self-pulsing can also occur when only half of the atoms in the cavity are driven, in this case the atom-cavity coupling phase can be identical for all atoms.

\emph{Conclusions.} -- %
We have proposed and studied a new variant of cavity enhanced Ramsey spectroscopy which simplifies and accelerates the measurement procedure via tailored atom-field coupling. As it is particularly advantageous for long lived atoms it should be beneficial for the performance of optical atomic clocks. The underlying phenomena are, on the one hand, the well-known cavity superradiance in an inverted atomic ensemble in order to achieve a fast readout, and, on the other hand, the subradiant behavior of specific atomic ensembles with an overall vanishing collective cavity coupling. Furthermore, we found that the chosen operating conditions with weak single atom coupling but strong collective coupling also induce a striking self-pulsing instability for continuous drive at suitable Rabi frequencies. Interestingly the necessary operating conditions are within the reach of current experimental setups with only minimal adjustments required.

Some preliminary studies on the influence of imperfections in the setup as variable coupling strengths, slow atomic motion or fluctuations in the excitation procedure qualitatively yield very similar results for experimentally realistic assumptions. However, a more detailed study of these and other aspects as heating and loss is required for a quantitative prediction of the practical system performance. 

\begin{acknowledgments}
We thank Stefan Sch\"affer for helpful discussions. We acknowledge funding from the European Union's Horizon 2020 research and innovation program under Grant Agreement No. 820404 iqClock. The numerical simulations were performed with the open-source framework QuantumCumulants.jl~\cite{plankensteiner2021quantumcumulantsjl} and  QuantumOptics.jl~\cite{kramer2018quantumoptics}.
\end{acknowledgments}

\bibliography{cavity-ramsey-method}

\clearpage

\appendix

\section{Mean-field equations}
Throughout the paper we calculate the dynamics with a second order cumulant expansion \cite{plankensteiner2021quantumcumulantsjl}. Nevertheless, the mean-field equations also contain the key physics, therefore we show these much simpler equations for a qualitative description of the system:
\begin{align}
\label{eq:meanfield_a}
\frac{d}{dt} \langle a\rangle  =& - \left( i \delta_\mathrm{c} + \frac{\kappa}{2} \right) \langle a\rangle  - i \sum_{j=1}^N g_j \langle {\sigma_j}^{{12}}\rangle
\end{align}
\begin{align}
\label{eq:meanfield_s22}
\frac{d}{dt} \langle {\sigma_j}^{{22}}\rangle  =& - \Gamma \langle {\sigma_j}^{{22}}\rangle  + i \frac{\Omega}{2} \left[ \langle {\sigma_j}^{{12}}\rangle - \langle {\sigma_j}^{{21}}\rangle \right] \\
+& i g_j \left[ \langle a^\dagger \rangle \langle {\sigma_j}^{{12}}\rangle - \langle a\rangle  \langle {\sigma_j}^{{21}}\rangle \right] \nonumber
\end{align}
\begin{align}
\label{eq:meanfield_s12}
\frac{d}{dt} \langle {\sigma_j}^{{12}}\rangle  =& \left( i \delta_\mathrm{a} - \frac{\Gamma + \nu}{2} \right) \langle {\sigma_j}^{{12}}\rangle \\
 +& i \left( \frac{\Omega}{2} + g_j \langle a\rangle \right) \left[ 2 \langle {\sigma_j}^{{22}}\rangle  - 1 \right] \nonumber
\end{align}

\begin{figure}[ht]
\centering
\includegraphics[width=\columnwidth]{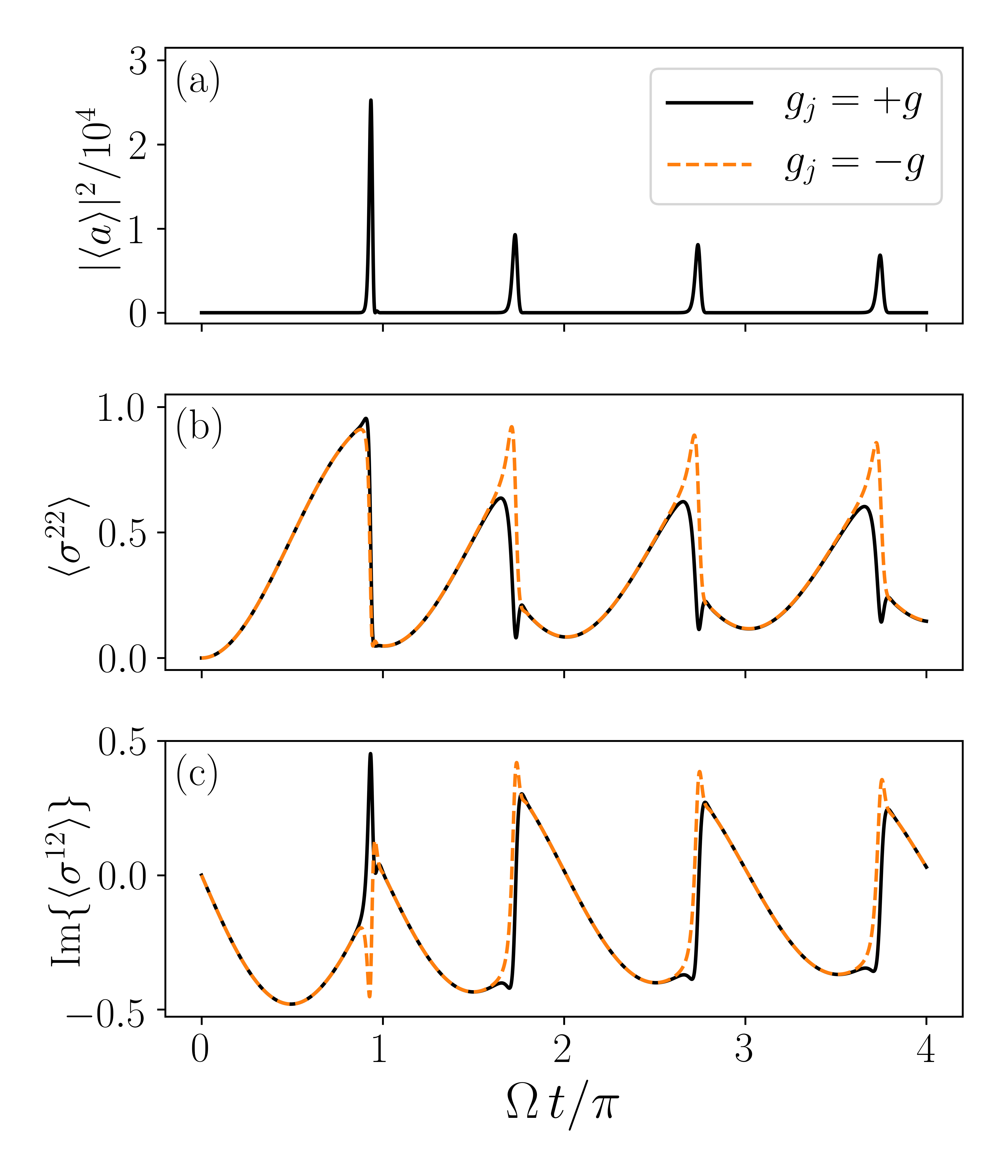}
\caption{\emph{Mean-field self-pulsing.} The timeevolution of the photon number (a) excited state population (b) and coherence (c) for a continuous drive. Due to the initial seed the behavior of the atoms in the two ensembles ($g_j = +g$ and $g_j = -g$) is striking different. The parameters are the same as in \fref{fig:dynamic_pulses} (a).}
\label{fig:meanfield_dynamic}
\end{figure}

In equation \eqref{eq:meanfield_a} we see that for a vanishing coupling phase and equal coherence for all atoms, the cavity field does not grow. We also observe this in \fref{fig:meanfield_dynamic}. On some point, however, the atoms will gain a sufficient inversion to emit a superradiant pulse into the cavity. Note that there is an initial inaccuracy (e.g. in the cavity field) needed in the mean-field description, otherwise the cavity photon number will never be unequal zero. Due to this initial inaccuracy the timeevolution for the two ensembles is striking different. For example we see that for the second pulse the excited state population is drastically decreased for one ensemble, but increased for the other. This is a feature of the mean-field description, which does not occur in that way in a higher order cumulant expansion.

Adiabatically eliminating the cavity field shows that the two frequencies $\Omega$ and $N g^2/\kappa$ compete with each other. Furthermore we see that for a small cavity field the atoms basically do independent Rabi-oscillations with frequency $\Omega$.

\section{Comparison with full quantum model}
\begin{figure*}[t]
\centering
\includegraphics[width=\textwidth]{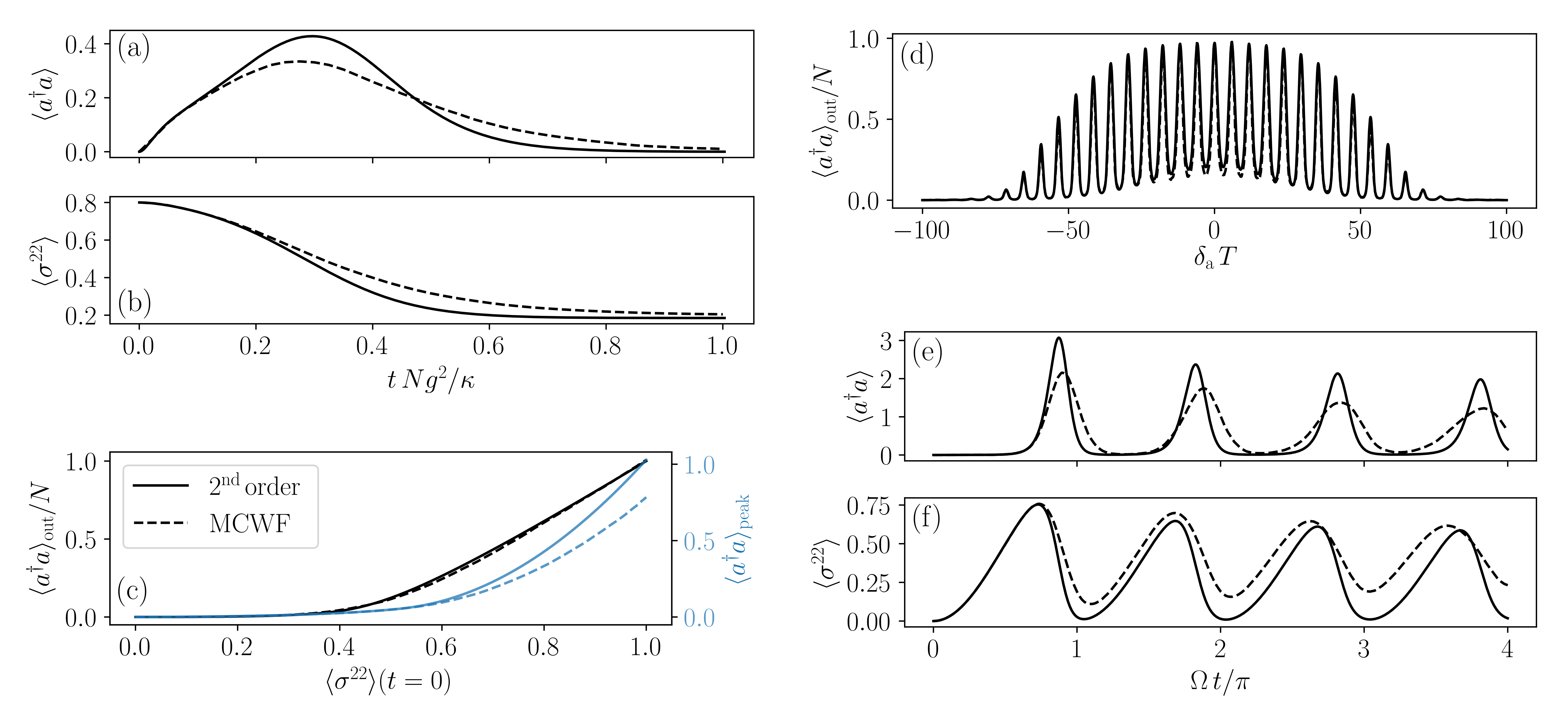}
\vspace{-0.6cm}
\caption{\emph{Full quantum model.} The $2^\mathrm{nd}$ order cumulant expansion (solid line) is compared with a full quantum model (dashed line) for the cavity subradiance in (a)-(c), the cavity Ramsey method in (d) and the self-pulsing in (e) and (f). In all plots we used $\kappa=200, \Gamma= \nu = \eta = 0$ and $\delta_\mathrm{c} = \delta_\mathrm{a}$. For (a)-(c) the remaining parameters are $N = 20, \, g=10$ and $\delta_\mathrm{a} = 0$, for (d) $N=20, \, g=10, \, \Omega = 100$ and $T=\pi / 10$ and for (e)-(f) $N = 2 \cdot 50, \, g=4, \, \delta_\mathrm{a} = 0$ and $\Omega = 4$.}
\label{fig:comparison}
\end{figure*}

To ensure the qualitative validity of our second order cumulant expansion we compare the results with a full quantum model. Of course this is only possible for a relative small number of atoms. To push the number of atoms as far as possible we use the Monte Carlo wave-function method \cite{dum1992monte, molmer1993monte}, and describe the atoms in the Dicke basis which means that only collective atomic effects can be investigated. Thus we neglect individual atomic decay and dephasing. \fref{fig:comparison} shows the comparison between the second order cumulant expansion and the full quantum model for the cavity subradiance (a)-(c), cavity Ramsey method (d) and self-pulsing (e)-(f).
Overall we find a good qualitative agreement. A perfect quantitative correspondence is not to be expected for such small atom numbers, due to the needed strong single atom cooperativity.

\end{document}